\documentclass[aps,prb,twocolumn,superscriptaddress, longbibliography]{revtex4-2}
\usepackage[utf8]{inputenc}
\usepackage[T1]{fontenc}
\usepackage{graphicx}
\usepackage{adjustbox}
\usepackage{subcaption}
\usepackage{dcolumn}
\usepackage{bm}
\usepackage{amsfonts}
\usepackage{amsmath}
\usepackage{amssymb}
\usepackage{color}
\usepackage[percent]{overpic}
\usepackage[normalem]{ulem}
\usepackage{soul}
\usepackage[colorlinks=true,citecolor=blue]{hyperref}
\hypersetup{colorlinks=true,citecolor=blue,linkcolor=red,urlcolor=blue}
\usepackage{ragged2e}
\makeatletter
\AtBeginDocument{
  \long\def\@makecaption#1#2{
    \par\smallskip
    \noindent
    \footnotesize
    \RaggedRight % <-- left aligned, NOT centered
    \justifying  % <-- fully-justified equal-length lines
    #1.\ #2\par
  }
}
\makeatother
\begin{document}

\title{Exploring the conventional and anomalous Josephson effects at arbitrary disorder strength in systems with spin-dependent fields}

\author{Maryam Darvishi}
\affiliation{Department of Physics and Nanoscience Center, University of Jyväskylä,
P.O. Box 35 (YFL), FI-40014 University of Jyv\"askyl\"a, Finland}

\author{F. Sebasti\'{a}n Bergeret}
\affiliation{Centro de F\'isica de Materiales (CFM-MPC) Centro Mixto CSIC-UPV/EHU, E-20018 Donostia-San Sebasti\'an,  Spain}
\affiliation{Donostia International Physics Center (DIPC), 20018 Donostia--San Sebasti\'an, Spain}

\author{Stefan Ili\'{c}}
\affiliation{Department of Physics and Nanoscience Center, University of Jyväskylä,
P.O. Box 35 (YFL), FI-40014 University of Jyväskylä, Finland}

\date{\today}

\begin{abstract}
We present a theory of the Josephson current in superconductor-normal metal-superconductor (SNS) junctions in the presence of generic spin-dependent fields, such as spin-orbit coupling (SOC), Zeeman fields, and altermagnetism.  We consider systems with arbitrary disorder strength, going beyond the usual diffusive and ballistic approximations. Using the linearized quasiclassical Eilenberger equation, we derive a compact expression for the Josephson current, which is then applied to various situations of experimental interest. First, we investigate the evolution of the Josephson critical current in an applied magnetic field in the presence of Rashba and Dresselhaus SOC, and discuss how this dependence can be used to probe SOC in the junction. We then study the anomalous Josephson ($\varphi_0$) effect in systems with Rashba SOC and show that it remains robust over a wide range of disorder strength, and can even be enhanced by moderate disorder in sufficiently long junctions. Finally, we investigate the Josephson current in disordered junctions with altermagnets, and show how the $0$-$\pi$ transition in such systems is suppressed by disorder. Our results may be useful for describing experimental setups with high-mobility samples, which nevertheless always contain some amount of disorder, and where neither purely ballistic nor diffusive approximations are adequate.
\end{abstract}

\maketitle

\section{Introduction}

The interplay between superconductivity and spin-dependent fields lies at the heart of many interesting phenomena explored in superconducting spintronics \cite{linder2015superconducting}. In this context, spin-dependent fields include magnetism as well as spin-orbit coupling (SOC). Their presence breaks two fundamental symmetries of the superconducting state, namely time-reversal and spatial inversion symmetries, respectively. As a consequence, the internal structure of Cooper pairs is modified, allowing spin-triplet and finite-momentum correlations to emerge \cite{bauer2012non}. This leads to distinct signatures in superconducting transport and enables new mechanisms for controlling dissipationless currents in spintronic devices. 

The simplest example of this physics arises in superconducting systems with a homogeneous Zeeman field \cite{buzdin2005proximity, bergeret2005odd}. In bulk superconductors, the competition of pairing and spin-splitting gives rise to Fulde-Ferrell-Larkin-Ovchinnikov (FFLO) states, where the Cooper pairs acquire a finite momentum and the order parameter becomes spatially modulated. In Josephson junctions, the same mechanism leads to oscillations of the superconducting condensate inside the ferromagnetic weak link. As a result, the critical current can change sign, producing the transition between the $0$ and $\pi$ states of the junction.  This $0$-$\pi$ transition is one of the main hallmarks of superconductor-ferromagnet Josephson junctions, and has been observed experimentally through variation of sample thickness and magnetic field.

Adding SOC to the picture modifies the Zeeman-field-driven physics described above \cite{amundsen2024colloquium}. For example, in Josephson junctions, the $0$-$\pi$ transition may be shifted to different fields or completely suppressed depending on the type of SOC and its strength. SOC may also enable the formation of equal-spin triplet correlations and the associated long-range proximity effect, in which superconducting correlations may persist over much longer distances than in the absence of such triplet components \cite{bergeret2013singlet,
bergeret2014spin}. Moreover, SOC enables an entirely new class of phenomena, known as magnetoelectric effects, in which magnetization and supercurrents become coupled \cite{bobkova2022magnetoelectric}. In bulk systems, magnetoelectric effects manifest as helical superconductivity \cite{agterberg2007magnetic, dimitrova2007theory} and Edelstein effects \cite{edelstein1995magnetoelectric,edelstein2005magnetoelectric}. In Josephson junctions, this mechanism produces an anomalous $\varphi_0$ phase shift corresponding to a finite supercurrent at zero phase difference \cite{buzdin2008direct, bergeret2015theory}. Perhaps the most striking manifestation of magnetoelectric effects is non-reciprocal transport in the form of the superconducting diode effect \cite{nadeem2023superconducting}, characterized by different critical currents for opposite current directions, which has recently attracted considerable attention and appears both in bulk systems and Josephson junctions. 

Lately, the interplay between altermagnetism and superconductivity has emerged as another active direction in superconducting spintronics \cite{sun2023andreev, papaj2023andreev, beenakker2023phase, ouassou2023dc, zhang2024finite, lu2024varphi, giil2024quasiclassical, kokkeler2025quantum}. Although experimental confirmation of the proximity effect in altermagnet-superconductor structures is still lacking, there have been a number of exciting theoretical developments. An altermagnet may be viewed as a system with momentum-dependent Zeeman splitting that averages to zero over the Brillouin zone \cite{vsmejkal2022emerging}. Despite the absence of net magnetization, coupling to superconductivity gives rise to similar phenomena as in magnetic systems, including FFLO-like modulation of the order parameter in the bulk \cite{zhang2024finite, chakraborty2024zero}, $0$-$\pi$ transitions in Josephson junctions \cite{ zhang2024finite, lu2024varphi}, along with new phenomena such as anisotropic current-phase relations \cite{ouassou2023dc}. 

Most existing theoretical descriptions of Josephson junctions in the presence of spin-dependent fields are formulated either in the ballistic or in the diffusive limit, corresponding respectively to negligible and strong impurity scattering, with only limited results available beyond these limits \cite{bergeret2001josephson, bergeret2007scattering}. Experimentally available structures, however, inevitably contain a finite amount of disorder, while modern material growth allows one to approach very high mobilities where the purely diffusive limit is no longer applicable. A theory valid at intermediate disorder strength is therefore required.

In this work, we develop such a theory within the quasiclassical formalism, describing the Josephson effect valid for arbitrary disorder strength in the presence of generic spin-dependent fields. The main focus of our work is the Josephson current in such junctions, given by the current-phase relation
\begin{equation}
j_J=j_s\sin\varphi+j_c\cos\varphi=|j_J|\sin(\varphi+\varphi_0),
\label{Eq:CPR}
\end{equation}
where $\varphi$ is the phase difference across the junction, and $j_s$ and $j_c$ are the conventional and anomalous \cite{buzdin2008direct, bergeret2015theory} Josephson currents, respectively. Alternatively, the current-phase relation can be written using the amplitude $|j_J|= \sqrt{j_{s}^2 + j_{c}^2}$ and the anomalous phase shift $\varphi_0 = \arctan(j_c/j_s)$. The conventional component $j_s$ occurs in all Josephson junctions, while the anomalous component $j_c$ and the associated phase shift $\varphi_0$ originate from magnetoelectric effects \cite{kokkeler2024nonreciprocal}. Their appearance requires simultaneous breaking of time-reversal symmetry and inversion symmetry, which can occur, for example, in the presence of SOC together with Zeeman fields. In this work, we investigate how the interplay of spin-dependent fields and disorder affects both $j_s$ and $j_c$. In Sec.~\ref{SecII}, we introduce the model and specify the geometry of the superconductor--normal metal--superconductor (SNS) Josephson junction and outline the quasiclassical Eilenberger formalism used throughout this work. 
In Sec.~\ref{SecIII}, we present the general solution of the Eilenberger equation, which provides a framework for calculating the Josephson current in the presence of generic spin-dependent fields for any disorder strength.

Our main findings are presented in Secs.~\ref{SecIV} and \ref{Sec:altermagnet}, where we apply our general theory to several systems of experimental interest. We first discuss junctions with Rashba and Dresselhaus SOC in the presence of a Zeeman field in Sec.~\ref{SecIVa}, where we demonstrate how the magnetic-field dependence of the critical current $j_s$ can be used to probe the nature and magnitude of the SOC. We then turn to the anomalous $\varphi_0$ effect in systems with Rashba SOC in Sec.~\ref{SecIVb}, and explore it over the full range of disorder strength. We demonstrate that the effect can even be enhanced by moderate disorder in sufficiently long junctions. Finally, we consider junctions incorporating altermagnetic materials in Sec.~\ref{Sec:altermagnet}, where we show that disorder leads to a rapid suppression of the $0$-$\pi$ transition. 

We summarize our main results and conclusions in Sec.~\ref{SecVI}.

\section{Model and setup \label{SecII}}
We consider an SNS Josephson junction where the normal region is subject to spin-dependent fields. The single-particle Hamiltonian describing the normal metal is
\begin{equation}
   H=\xi_\mathbf{p}+b_{{\mathbf{p}}}^i \sigma_i+h_{{\mathbf{p}}}^i \sigma_i+V_{imp}.
   \label{Eq:normal}
\end{equation}

Here and throughout the text, we use the natural units where $k_B=\hbar=1$, $\xi_\mathbf{p} = \frac{\mathbf{p}^2}{2m} - \mu$ is the kinetic energy term, where $ \mu$ is the chemical potential and $\mathbf{p}$ is the momentum. Spin-dependent fields are captured by the SOC terms  $b_\mathbf{p}^i$, and the magnetic (Zeeman-like) terms $h_\mathbf{p}^i$. SOC terms break the spatial inversion symmetry and are odd under momentum inversion, $b_{{-\mathbf{p}}}^i=-b_{{\mathbf{p}}}^i$, whereas magnetic terms break the time-reversal symmetry (TRS) and are even under momentum inversion $h_{{-\mathbf{p}}}^i=h_{{\mathbf{p}}}^i$. We introduced the Pauli matrices $\sigma_{i}$, $i=x,y,z$,  acting in spin space. The potential $V_{imp}$ represents a random spin-independent impurity potential. Throughout this work, we will consider only linear-in-momentum SOC, so that $b_\mathbf{p}^i=b^{ij}p_j$, and we will retain momentum dependence in the magnetic terms  $h_\mathbf{p}^i$, which will be essential for the description of altermagnetism. 

\begin{figure}[htbp]
  {\centering
    \includegraphics[width=0.4\textwidth]{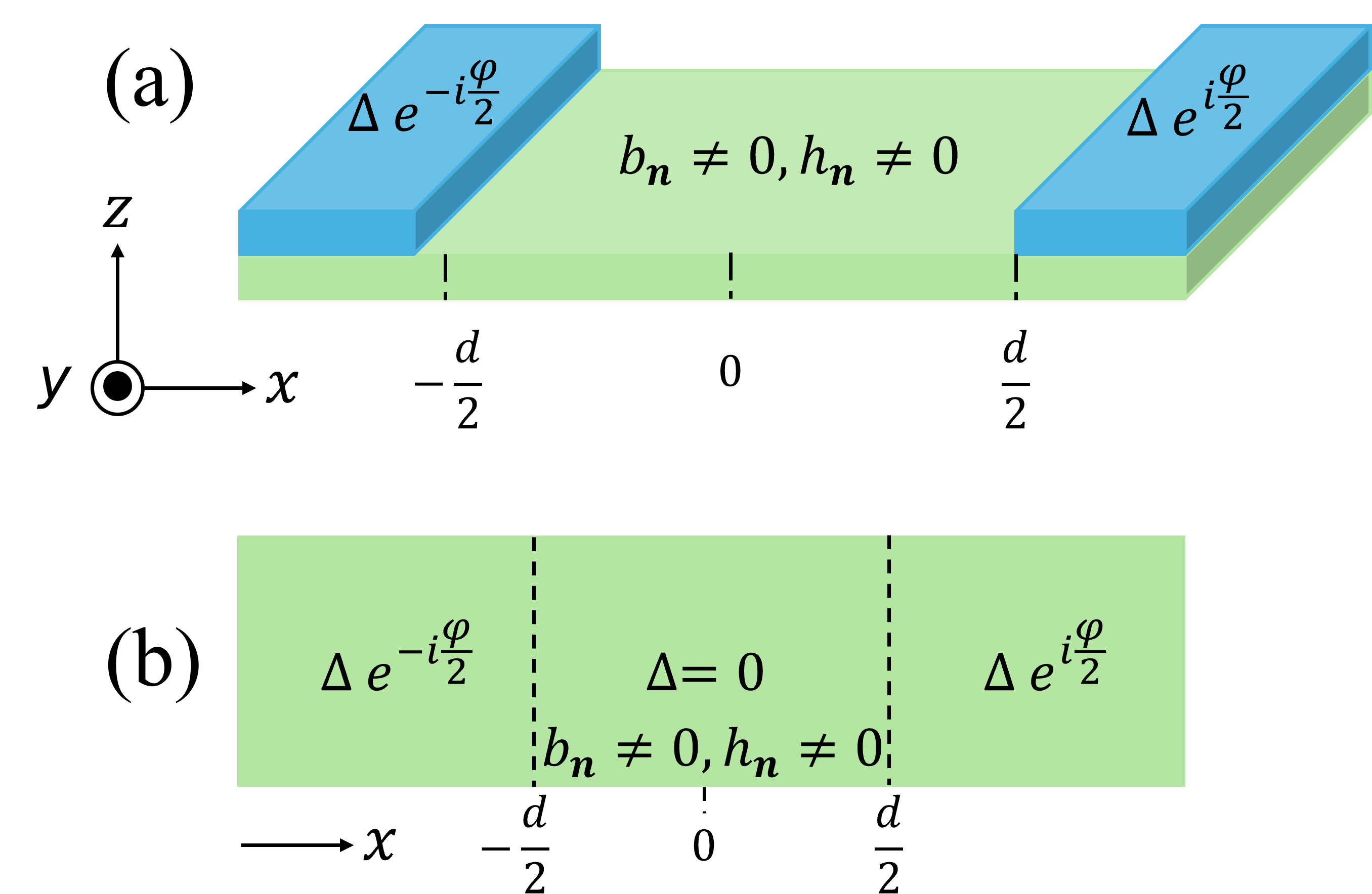}\par
  }
  \caption{Schematic of an SNS junction with spin-dependent fields. 
  (a) Typical experimental setup: two s-wave superconducting leads are deposited on top of a normal metal, forming a planar SNS Josephson junction.
  (b) Theoretical model used in this work: an infinite 2D material with spin-dependent fields in which a finite pairing potential is imposed at $|x|>d/2$.  
  Spin-dependent fields are present throughout the entire structure, including both the regions where $\Delta=0$ and $\Delta\neq 0$.}
  \label{model}
\end{figure}

Figure~\ref{model} shows the system under consideration. Panel \ref{model}(a) illustrates a typical experimental configuration: a planar Josephson junction in which two large superconducting electrodes are deposited on a normal metal that hosts spin-dependent fields. In our theoretical treatment, this structure is approximated by considering an infinite two-dimensional system where a conventional $s$-wave superconducting pairing potential is imposed in the regions $|x|>d/2$, as shown in Fig.~\ref{model}(b).  
Therefore, the pairing potential has the form
\begin{equation}
\Delta(x) = \Delta e^{i\frac{\varphi}{2}} \Theta\left(x - d/2\right) + \Delta e^{-i\frac{\varphi}{2}} \Theta\left(-x - d/2\right),
\label{Eq:pairing}
\end{equation}
where $\Delta$ is its magnitude, $d$ denotes the distance between the superconducting electrodes, $\varphi$ is the phase difference between the two superconductors, and $\Theta$ is the Heaviside step function. Spin-dependent fields are taken to be present throughout the entire structure, including both the regions with $\Delta=0$ and $\Delta\neq 0$. 

Combining the normal state Hamiltonian \eqref{Eq:normal} with the pairing potential \eqref{Eq:pairing}, we can write the Bogoliubov-de Gennes Hamiltonian for our setup
\begin{equation}
    H_{\rm BdG}=[\xi_\mathbf{p}+b_{{\mathbf{p}}}^i\sigma_i +V_{imp}]\tau_{z}+h_{{\mathbf{p}}}^i\sigma_i+\Delta(x)\tau_{+}+\Delta^{*}(x)\tau_{-},
\end{equation}
where $\tau_{i}$ are Pauli matrices in Nambu space, and $\tau_{\pm}=\frac{1}{2}(\tau_{x}\pm i\tau_{y}).$
Next, we assume the quasiclassical approximation, meaning the chemical potential is the dominant energy scale $\mu\gg \Delta, \tau^{-1}, h_\mathbf{p}^i, b_\mathbf{p}^i$, and that electronic transport occurs near the Fermi level. This assumption holds in the majority of experimentally available materials, and moreover, it enables us to approximate the spin dependent fields as $b_\mathbf{p}^i\approx b_{p_F \mathbf{n}}^i\equiv b_{\mathbf{n}}^i$ and   $h_\mathbf{p}^i\approx h_{p_F \mathbf{n}}^i\equiv h_{\mathbf{n}}^i$, with $p_F$ being the Fermi momentum, and $\mathbf{n}=\mathbf{p}/|\mathbf{p}|$ describes the momentum direction along the Fermi surface. Furthermore, we assume that superconducting correlations $\Delta$ are weak,  which is appropriate for structures in which $\Delta$ is introduced via the proximity effect through an opaque superconductor-normal interface, or close to the critical temperature of the superconductor.

Under these assumptions, our system is described by the linearized Eilenberger equation \cite{bergeret2014spin,bergeret2015theory,konschelle2015theory}
\begin{multline}
    v_{F} n_{k} \partial_{k}\hat{f} + 2 \omega_n \hat{f} - i\{h_\mathbf{n}^i\sigma_i, \hat{f}\} + i [b_{{\mathbf{n}}}^i\sigma_i , \hat{f}] \\ =\frac{1}{\tau}(\langle\hat{f}\rangle - \hat{f})  - 2i \Delta(x)+ \frac{1}{2m}  \{\hat{\mathcal{F}}_{ij}, n_{i} \partial_{n_j} \hat{f}\}.
    \label{Eq:Eilenberger}
\end{multline}
Here, $\hat{f}=f^i \sigma_i$, $(i=0,x,y,z)$ is quasiclassical anomalous Green’s function, where $f^0$ describes the singlet superconducting correlations, while $f^{x,y,z}$ describes the triplets. Furthermore,  $\omega_{n}=2 \pi T (n+1/2)$, $n\in Z$,  is the Matsubara frequency, with $T$ being the temperature, $\tau$ is the disorder scattering time, and $\langle...\rangle$ denotes averaging over the momentum-direction $\mathbf{n}$. As seen from the structure of Eq.~\eqref{Eq:Eilenberger}, the main role of the magnetic terms $h_\mathbf{n}^i$ is to couple singlets and triplets. The role of the SOC field is two-fold. First, the term $[b_{{\mathbf{n}}}^i\sigma_i, \hat{f}]$ introduces coupling between triplets, leading to triplet precession and relaxation. Second, the term containing $\hat{\mathcal{F}}_{ij}$ is responsible for magnetoelectric phenomena, such as the anomalous Josephson effect, the Edelstein effects, and the diode effect. The object $\hat{\mathcal{F}}_{ij}=-\hat{\mathcal{F}}_{ji}=-i[\hat{\mathcal{A}}_i,\hat{\mathcal{A}_j}]$ is known as the field-strength tensor, with $\hat{\mathcal{A}}_i=-2 b^{ki}\sigma_k$  corresponding to the effective vector potentials in  the SU(2) gauge-field formulation of linear-in-momentum spin–orbit coupling \cite{tokatly2008equilibrium, gorini2010non, bergeret2014spin}. Importantly, this magnetoelectric term scales as $\sim |b_\mathbf{n}|/\mu$, and therefore it is necessarily small within the quasiclassical approximation. Still, we retain it to be able to discuss the anomalous Josephson effect in Sec.~\ref{SecIVb}.

In terms of the anomalous Green's functions, the supercurrent is given by:
\begin{equation}
   j_{i}= -i e \pi N_{0} v_{F}T \sum_{\omega_n>0} \text{Tr}_{\sigma}\langle n_{i} \hat{f}\hat{\bar{f}}\rangle,
   \label{eq:current}
\end{equation}
where $N_{0}$ is the density of states at the Fermi level and $\text{Tr}_{\sigma}$ denotes the trace over spin, and $\hat{\bar{f}}
(\mathbf{n})=\sigma_{y}\hat{f}^{*}(\mathbf{-n})\sigma_{y}$ is the time-reversal conjugated anomalous Green's function. This current is conserved, $\nabla \mathbf{j}=0$, provided that $\Delta(x)$ is determined self-consistently \cite{konschelle2015theory}. For simplicity, and to avoid the complexity of the self-consistent calculation, we assume that $\Delta$ is rigid and unaffected by the weak link and the spin-dependent fields.  This assumption is justified in lateral junctions shown in Fig.~\ref{model}(a), where superconductivity is induced in a 2D normal metal by large 3D superconducting electrodes.

In the next section, we present a general solution of Eq.~\eqref{Eq:Eilenberger}. This solution is then used to calculate the Josephson current using Eq.~\eqref{eq:current}.

\section{General Solution of the Eilenberger equation and the Josephson current\label{SecIII}}

To solve Eq.~\eqref{Eq:Eilenberger}, we first notice that the spatial variations in our setup occur only along the $x$-direction, meaning that the problem is effectively one-dimensional. We proceed by transforming Eq.~\eqref{Eq:Eilenberger} to the reciprocal space, by applying the Fourier transform
\begin{equation}
F(x)=\frac{1}{\sqrt{2 \pi}} \int d Q e^{i Q x} F_Q \tag{12}.
\label{eq:fourier}
\end{equation}
This way we obtain
\begin{multline}
 \left(i Q v_F n_x+2 \omega_n+\tau^{-1}\right) \hat{f}_Q-i\{ h_{{\mathbf{n}}}^i\sigma_i,\hat{f}_Q\}+i[ b_{{\mathbf{n}}}^i\sigma_i, \hat{f}_Q]\\ =\tau^{-1}\langle \hat{f}_Q\rangle+\frac{1}{2m}\{\hat{\mathcal{F}}_{xy},\mathcal{L} \hat{f}_Q\}+{{\Delta}}_Q,
 \end{multline}
where we introduced the shorthand notation for the operator $\mathcal{L}\equiv n_x \partial_{n_y}-n_y \partial_{n_x}$, and the source term $\Delta_Q$ is obtained by Fourier-transforming the pair potential $\Delta(x)$,
\begin{multline}
    \Delta_Q = -2i\Delta \sqrt{2\pi} \delta(Q) \cos\left(\varphi/2\right) \\
    + \frac{2\Delta}{\sqrt{2\pi} Q} 
    \left[ e^{i d(Q + \varphi)/2} - e^{-i d(Q + \varphi)/2} \right].
    \label{eq:source_term}
\end{multline}
Next, we express the anomalous Green's function as a four-component vector in the singlet–triplet basis:
\begin{equation}
    \vec{f}_Q = (f_{Q}^x, f_{Q}^y, f_Q^z,f_Q^0)^\mathrm{T},
    \label{eq:vector_notation}
\end{equation} 
Now, we can rewrite the Eilenberger equation in the following form 
\begin{equation}
\left[\widehat{\Pi}^{(0)}\right]^{-1}\vec{f}_Q=\frac{1}{\tau} \langle \vec{f}_Q\rangle+  \widehat{\mathcal{F}} \mathcal{L}\vec{f}_Q+\Delta_Q \vec{e}_0,
\label{Eq:vecEilenberger}
\end{equation}
where $\vec{e}_0 = (0, 0, 0, 1)^\mathrm{T}$. Here we introduce the polarization operator $\widehat{\Pi}^{(0)}$, which is a $4\times4$ operator acting in a singlet-triplet space defined as
\begin{equation}
    \widehat{\Pi}^{(0)} = \left(\tau^{-1} + 2\omega_n + iQ v_F n_x + \widehat{\mathcal{A}}_\mathbf{n} \right)^{-1},
    \label{eq:pi0}
\end{equation}
where \( \widehat{\mathcal{A}}_\mathbf{n} \) is a matrix that accounts for spin-dependent fields. 
\begin{equation}
\widehat{\mathcal{A}}_\mathbf{n} = 2 \begin{pmatrix}
0 & b_\mathbf{n}^z & -b_{\mathbf{n}}^y & -i h^x_\mathbf{n} \\
-b^z_\mathbf{n} & 0 & b^x_\mathbf{n} & -i h^y_\mathbf{n} \\
b^y_\mathbf{n} & -b^x_\mathbf{n} & 0 & -i h^z_\mathbf{n} \\
-i h^x_\mathbf{n} & -i h^y_\mathbf{n} & -i h^z_\mathbf{n} & 0
\end{pmatrix}.
\label{eq:A_matrix}
\end{equation}
Magnetoelectric effects are captured by the operator $\widehat{\mathcal{F}}$ given as
\begin{equation}
  \widehat{\mathcal{F}} = \begin{pmatrix}
0 & 0 & 0 & \mathcal{F}_x \\
0 & 0 & 0 & \mathcal{F}_y \\
0 & 0 & 0 & \mathcal{F}_z \\
\mathcal{F}_x & \mathcal{F}_y & \mathcal{F}_z & 0
\end{pmatrix},
\label{eq:F_tensor}
\end{equation}
where $\mathcal{F}_i= \frac{1}{4m}\text{Tr} \left(\sigma_i \hat{\mathcal{F}}_{xy}\right)$.

We proceed to solve Eq.~\eqref{Eq:vecEilenberger} perturbatively in the magnetoelectric term $\widehat{\mathcal{F}}$. As explained in Sec.~\ref{SecII}, such a perturbative treatment is justified because magnetoelectric effects are weak in the quasiclassical theory. We expand the Green's function as $\vec{f}_Q = \vec{f}_Q^{(0)} + \delta \vec{f}_Q$, where $\vec{f}^{(0)}_Q$ is the bare solution, and $\delta \vec{f}_Q\propto \widehat{\mathcal{F}}$ is the perturbative correction. Using Eq.~\eqref{Eq:vecEilenberger}, we find 
\begin{equation}
    \vec{f}_Q^{(0)} = \frac{1}{\tau}\widehat{\Pi}^{(0)}\langle \vec{f}^{(0)}_Q\rangle+\widehat{\Pi}^{(0)} \Delta_Q \vec{e}_0,  
    \label{eq:f0}
\end{equation}
\begin{equation}
    \delta \vec{f}_Q = \frac{1}{\tau}\widehat{\Pi}^{(0)}\langle \delta \vec{f}_Q\rangle+\widehat{\Pi}^{(0)} \widehat{\mathcal{F}} \mathcal{L}\vec{f}^{(0)}_Q.
    \label{eq:df}
\end{equation}
Solving these equations finally yields
\begin{equation}
    \vec{f}_Q = \widehat{\Pi} \left[1 - \frac{1}{\tau}\langle \widehat{\Pi}\rangle \right]^{-1} \Delta_Q \vec{e}_0,
    \label{eq:full_solution}
\end{equation}
where we introduced the total polarization operator as
\begin{equation}
    \widehat{\Pi} = \widehat{\Pi}^{(0)} + \delta \widehat{\Pi},
    \label{eq:pi}
\end{equation}
consisting of a "bare" contribution defined in Eq.~\eqref{eq:pi0} corrected to account for magnetoelectric effects by
\begin{align}
    \delta \widehat{\Pi} =  \widehat{\Pi}^{(0)} \widehat{\mathcal{F}} \widehat{\Pi}^{(0)}
    \mathcal{L} \left[\widehat{\Pi}^{(0)}\right]^{-1}
    \widehat{\Pi}^{(0)}.
    \label{eq:delta_pi}
\end{align}
Equation ~\eqref{eq:full_solution}, together with Eqs.~\eqref{eq:pi0}, \eqref{eq:pi} and \eqref{eq:delta_pi}, provides a generic solution for the anomalous Green's function in the reciprocal $Q$-space in the presence of arbitrary disorder strength and for generic spin-dependent fields. It is one of the main results of our work. We proceed to calculate the Josephson current in different cases based on these equations. 

\subsection{Calculation of the Josephson current \label{SecIIIa}}
To compute the supercurrent, we first need to transform the anomalous Green's function from the $Q$-space back to the real space. At the center of the junction ($x=0$), we obtain, using \eqref{eq:fourier}, $\vec{f}(0)=\frac{1}{\sqrt{2\pi}}\int dQ \vec{f}_Q$. In general, this integral must be evaluated numerically, as analytical results can be obtained only in the ballistic and diffusive limits, and even then, only for some specific configurations of spin-dependent fields, as discussed below. The Josephson current then follows from Eq.~\eqref{eq:current}. 
\subsection{Ballistic and diffusive limit \label{SecIIIb}}
In this section, we examine how our theory simplifies in the ballistic and diffusive limits. 
\paragraph{Ballistic limit.} In the absence of disorder, Eq.~\eqref{eq:full_solution} simplifies to
\begin{equation}
    \vec{f}_Q = \widehat{\Pi} \Delta_Q \vec{e}_0,
    \label{eq:ballistic}
\end{equation}
where one should set $\tau^{-1}\to 0$ when evaluating the polarization operator $\eqref{eq:pi0}$.
This form enables simpler numerical evaluation and, in certain cases, analytical calculations of the Josephson current. 

To illustrate how the current is obtained from Eq.~\eqref{eq:ballistic}, let us consider the simplest case -- the absence of spin-dependent fields. Then, only the singlet component remains,  $f_Q^0=\Delta_Q/(2\omega_n+i Q v_F n_x )$, which, after transforming to real space becomes $f^{0}(0)=\frac{\Delta}{\omega_n} \exp\left(-d\omega_n/|n_x|v_F\right)(\operatorname{sgn}(n_x)\sin \frac{\varphi}{2}-i\cos \frac{\varphi}{2})$. From here, we finally obtain the Josephson current 
\begin{equation}
j_J= 2 e N_{0} v_{F} \pi T \Delta^2 \sum_{\omega_n>0}\frac{1}{ \omega_n^2}\left \langle |n_x|
 e^{-\frac{2 d \omega_n}{v_F|n_x|}} \right \rangle \sin(\varphi).
 \end{equation}
 This expression is consistent with the previous works on ballistic Josephson junctions in the limit of small $\Delta$ \cite{ golubov2004current}. In Appendix \ref{AppA}, we present a similar analytical solution for a system with Rashba SOC and an in-plane Zeeman field, which is discussed in more detail in Sec.~\ref{SecIVa}. 
\paragraph{Diffusive limit.} At strong disorder, $\tau^{-1}\gg \Delta, h_\mathbf{n}^i, b_\mathbf{n}^i$, we may expand the polarization operator in orders of $\tau$ as $\widehat{\Pi}^{(0)}\approx \tau(1-\widehat{\Lambda}\tau+\widehat{\Lambda}^2\tau^2+...)$ with $\widehat{\Lambda}=2\omega_n+iQv_F n_x+\widehat{\mathcal{A}}_\mathbf{n}$. The leading orders in the magnetoelectric correction term are obtained as $\delta\Pi\approx \tau^3\widehat{\mathcal{F}}\mathcal{L}\widehat{\Lambda} -\tau^4 (\widehat{\mathcal{F}}\widehat{\Lambda}\mathcal{L}\widehat{\Lambda}+\widehat{\mathcal{F}}(\mathcal{L}\widehat{\Lambda})\widehat{\Lambda}+\widehat{\Lambda}\widehat{\mathcal{F}}\mathcal{L}\widehat{\Lambda})$. Then, we obtain the anomalous Green's function as 
\begin{multline}
\vec{f}_Q\approx \big(1-iQ v_F \tau n_x -\widehat{\mathcal{A}}_\mathbf{n} \tau+2iQv_F n_x \widehat{\mathcal{A}}_\mathbf{n}\tau^2
\\
+\widehat{\mathcal{F}} \mathbf{n}\mathcal{L}\widehat{\mathcal{A}}_\mathbf{n}\tau^2 \big) \widehat{\mathcal{D}} \Delta_Q \vec{e}_0.
\label{Eq:Difsol}
\end{multline}
In the parentheses in the above equation, one must retain terms that are odd in $\mathbf{n}$, since these terms are essential for obtaining a finite supercurrent. We introduced the diffusion operator
\begin{multline}
\widehat{\mathcal{D}}^{-1}\equiv \frac{1}{\tau}\left(1-\frac{1}{\tau}\langle\Pi\rangle\right)\approx 2\omega_n  + \langle \widehat{\mathcal{A}}_\mathbf{n} \rangle +D Q^2  -
\langle \widehat{\mathcal{A}}_\mathbf{n}^2\rangle \tau\\-2iQ v_{F} \langle n_x \widehat{\mathcal{A}}_\mathbf{n}\rangle\tau-6 D Q^2 \langle n_x^2 \widehat{\mathcal{A}}_\mathbf{n}\rangle\tau -iQ [\langle n_y \widehat{\mathcal{A}}_\mathbf{n} \rangle, \widehat{\mathcal{F}}]\tau^2.
\label{eq:diffusion}
\end{multline}
Here, $D=\frac{1}{2}v_F^2\tau$ is the diffusion constant. The physical role of different terms is the following. The term $\langle \widehat{\mathcal{A}}_\mathbf{n} \rangle$ describes the average spin splitting, to which only the magnetic fields $h_\mathbf{n}^{i}$ contribute.  On the other hand, only the SOC fields $b_\mathbf{n}^i$ contribute to the term $2iQ \langle n_x \widehat{\mathcal{A}}_\mathbf{n}\rangle\tau$, capturing triplet precession, that is, rotations of the triplet vector in real space caused by the SOC. The term $\langle \widehat{\mathcal{A}}_\mathbf{n}^2\rangle \tau$ governs the triplet and singlet relaxation due to the SOC and magnetic fields, respectively.  The term $6 D Q^2 \langle n_x^2 \widehat{\mathcal{A}}_\mathbf{n}\rangle\tau$ captures spin-dependent corrections to the diffusion constant due to magnetic fields. As we discuss in Sec.~\ref{Sec:altermagnet}, this term is needed to capture the $0$-$\pi$ transitions in diffusive altermagnets. Finally, the last term in Eq.~\ref{eq:diffusion} accounts for magnetoelectric effects, and it is necessary to obtain the anomalous Josephson effect, discussed in Sec.~\ref{SecIVb}. 

The solution presented in Eq.~\eqref{Eq:Difsol} is equivalent to solving the linearized Usadel equation for diffusive systems \cite{belzig1999quasiclassical,virtanen2022nonlinear,kokkeler2025universal}. As a specific example, for a system with Rashba SOC and an in-plane Zeeman field, Eq.~\eqref{eq:diffusion} reproduces the well-known diffusion operator \cite{burkov2004theory}, widely used in both normal and superconducting spintronics. Its explicit form is shown in Appendix \ref{AppB}.

Let us illustrate how to obtain the Josephson current using Eq.~\eqref{Eq:Difsol} for the simplest case with no spin-dependent fields.  In this case, the only finite component is the singlet one,  $f^0_Q=\Delta_Q (1- i Q v_F \tau n_x )/(D Q^2 +2\omega_n)$,
which, when transformed into real space, yields $f^0(0)=\frac{\Delta}{\omega_n} \exp(-d\kappa_\omega /2)(\kappa_\omega n_x v_F \tau \sin \frac{\varphi}{2}-i \cos \frac{\varphi}{2})$. Here we introduced $\kappa_\omega=\sqrt{2\omega_n/D}$. The corresponding Josephson current is then \cite{golubov2004current, strambini2020josephson}
\begin{equation}
j_J=2e N_{0} \pi T D\sum_{\omega_n>0}\frac{\Delta^2 \kappa_\omega e^{-d \kappa_\omega}}{\omega_n^2}\sin(\varphi) . 
\end{equation}

\section{Application to systems with SOC and Zeeman field \label{SecIV}}

In this section, we apply our general result to systems with linear-in-momentum SOC subjected to in-plane Zeeman fields. This setting is directly relevant to Josephson junctions formed by superconductors (e.g., Al) proximitizing semiconducting 2D electron gases with strong SOC (e.g., InAs) \cite{fornieri2019evidence, assouline2019spin, mayer2020gate, strambini2020josephson, dartiailh2021phase, baumgartner2022supercurrent, banerjee2023signatures, mandal2024magnetically, reinhardt2024link}, and also to junctions formed in van der Waals heterostructures that combine superconductivity with intrinsic SOC and magnetism \cite{yabuki2016supercurrent,ai2021van, idzuchi2021unconventional, wu2022field}. Such platforms are central in superconducting spintronics and have been recently extensively investigated in experiments \cite{amundsen2024colloquium}. A major recent focus in these systems has been on magnetoelectric phenomena, most notably the diode effect \cite{nadeem2023superconducting} and the anomalous $\varphi_0$ Josephson effect. Understanding how these phenomena depend on disorder is therefore both of fundamental and practical interest. 

We first neglect magnetoelectric effects in Sec.~\ref{SecIVa}, which means setting $\widehat{\mathcal{F}}\to 0$ in Eq.~\eqref{eq:full_solution}. Our goal is to first investigate how the interplay between SOC and Zeeman fields controls the conventional Josephson effect, described by $j_s$ in Eq.~\eqref{Eq:CPR}, specifically in systems with linear Rashba and Dresselhaus SOC. In particular, we discuss how the critical current exhibits characteristic dependence on the direction of in-plane field, which can be used to identify and probe the type of SOC present in the system. 

Then, in Sec.~\ref{SecIVb} we consider magnetoelectric phenomena, taking finite $\widehat{\mathcal{F}}$, and investigate the anomalous Josephson effect in systems with linear Rashba SOC. While this effect has been extensively studied in the ballistic \cite{buzdin2008direct, konschelle2015theory, hasan2022anomalous} and diffusive limits\cite{bergeret2015theory, ilic2024superconducting}, the intermediate disorder regime, which is likely most relevant for experiments, has remained largely unexplored. We analyze the full crossover from the clean to diffusive regime, and find that the anomalous phase shift can even be enhanced by moderate disorder in sufficiently long junctions.  

In all plots throughout this work, we set the temperature to $T = 0.1 \Delta$, and we express lengths in terms of the coherence length of a clean superconductor $\xi_0=v_F/\Delta$.

\subsection{Conventional Josephson effect and $0$-$\pi$ transitions in the presence of SOC \label{SecIVa}}

Spin-orbit coupling, on its own, does not affect the Josephson effect in conventional s-wave junctions. This changes once the Zeeman field is applied. Namely, the Zeeman field converts the singlet correlations into triplet ones, and SOC then couples the triplets and causes them to precess. As a consequence, SOC can significantly modify the $0$-$\pi$ transitions, shifting the field at which the transition happens or eliminating the transition altogether. Related effects have been studied in systems with spin-orbit impurities \cite{bergeret2007scattering}, and in junctions with intrinsic SOC in ballistic\cite{konschelle2016ballistic} and diffusive limits\cite{arjoranta2016intrinsic}. Here, we extend those works to arbitrary disorder strength. 

We consider a system with linear Rashba and Dresselhaus spin-orbit coupling subjected to a homogeneous in-plane Zeeman field. In this case, the corresponding SOC terms that should be substituted into Eq.~\eqref{eq:A_matrix} read
\begin{equation}
b_{\mathbf{n}}^x = \alpha p_F\, n_y + \beta p_F\, n_x, \, \quad
b_{\mathbf{n}}^y = -\alpha p_F\, n_x - \beta p_F\, n_y,
\label{Eq:coefb}
\end{equation}
where $\alpha$ and $\beta$ describe the Rashba and Dresselhaus SOC, respectively, while the Zeeman field is captured by
\begin{equation}
h_\mathbf{n}^x=h_x, \qquad h_\mathbf{n}^y=h_y.
\label{Eq:coefh}
\end{equation}

We first consider the case of pure Rashba SOC, $\alpha\neq 0$ with $\beta=0$. In the ballistic limit, the Josephson current can be obtained analytically, although the expression is somewhat cumbersome and deferred to Appendix \ref{AppA}. In the diffusive limit, the closed-form solution is no longer possible analytically, but the diffusion operator acquires a simple form known from previous studies \cite{burkov2004theory}, see Appendix \ref{AppB}. In Fig.~\ref{fig:SOCj_d}, we show how Rashba SOC modifies the critical current $j_s$. Without SOC (blue curve), the critical current changes sign as the junction length increases -- this is the $0$-$\pi$ transition. A similar sign change also occurs when the length of the junction is fixed, and the Zeeman field is varied. In the presence of Rashba SOC, this behavior becomes anisotropic. For a field applied along the $y$ direction, perpendicular to the current, the $0$-$\pi$ transition remains, but it is shifted [see Fig.~\ref{fig:SOCj_d}(a)]. In contrast, for a field applied along the $x$-direction, parallel to the current, the transition is suppressed at sufficiently strong SOC [see Fig.~\ref{fig:SOCj_d}(b)]. Namely, for quasiparticles moving along $x$ (the direction to the current), Rashba SOC causes spin precession around the $y$-axis. A Zeeman field applied along $x$ generates the $x$-triplet, which is strongly affected by the precession, which then ultimately disrupts the $0$-$\pi$ transition at sufficiently strong SOC. In contrast,  a field along $y$ produces a $y$-triplet that does not precess, allowing the $0$-$\pi$ transition to persist at strong SOC. 
\begin{figure}
    \centering
        \makebox[\linewidth][l]{\textbf{}}
        \includegraphics[width=\linewidth]{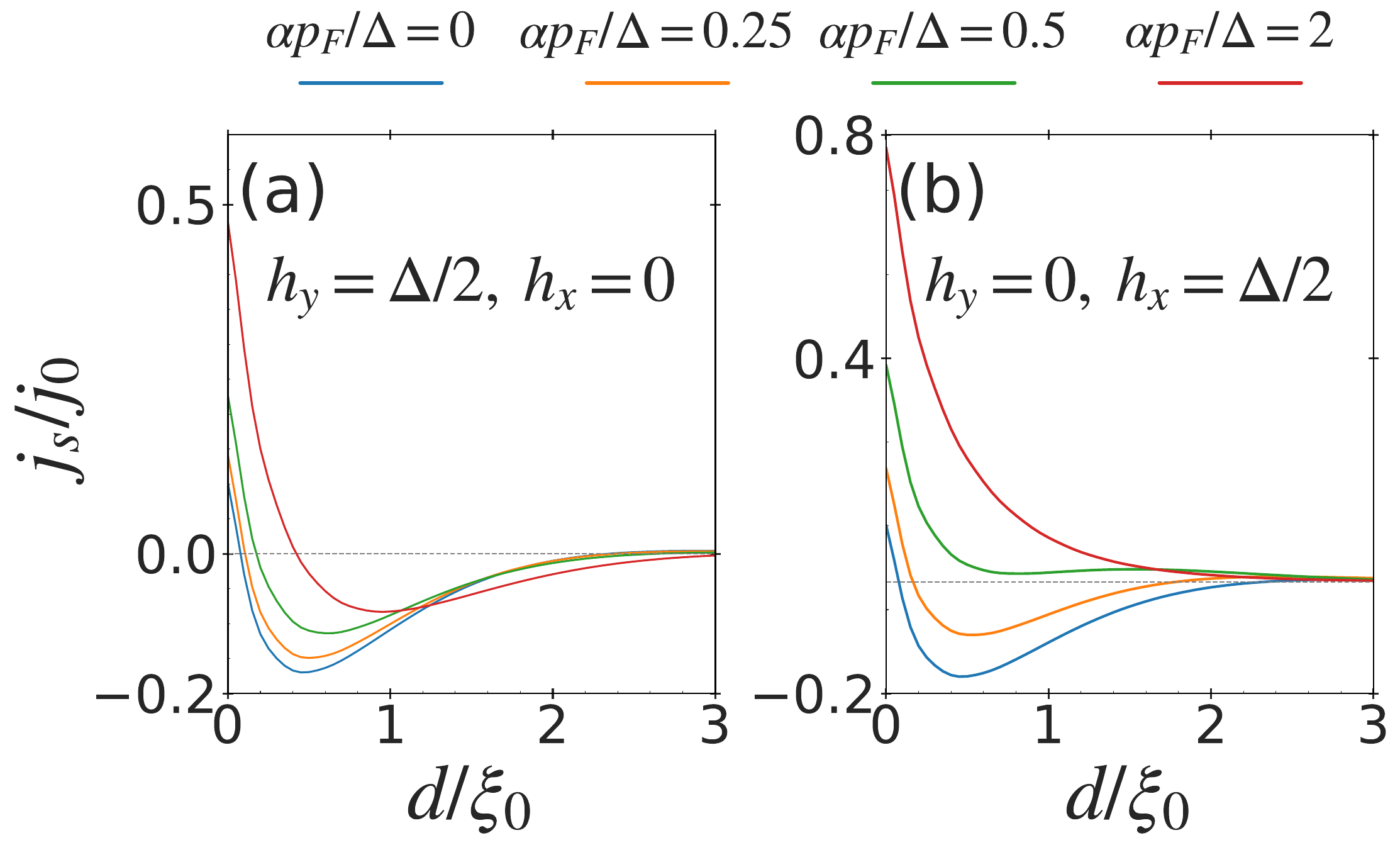} \\
    \caption{
        Josephson critical current as a function of junction length for several strengths of Rashba spin–orbit coupling. 
       (a) Zeeman field applied perpendicular to the current (\( h_y=\Delta/2,\, h_x=0 \)). 
        (b) Zeeman field applied parallel to the current (\( h_x=\Delta/2,\, h_y=0 \)). 
        In the absence of SOC, the current exhibits a \( 0 \)–\( \pi \) transition with increasing length, while finite Rashba coupling introduces a pronounced anisotropy and can suppress the transition for fields parallel to the current. 
        The current is normalized by \( j_0=2eN_0v_F \), and disorder is fixed at \( 1/\tau=\Delta \).
            }
    \label{fig:SOCj_d}
\end{figure}

\begin{figure}
    \centering
        \makebox[\linewidth][l]{\textbf{}}
        \includegraphics[width=\linewidth]{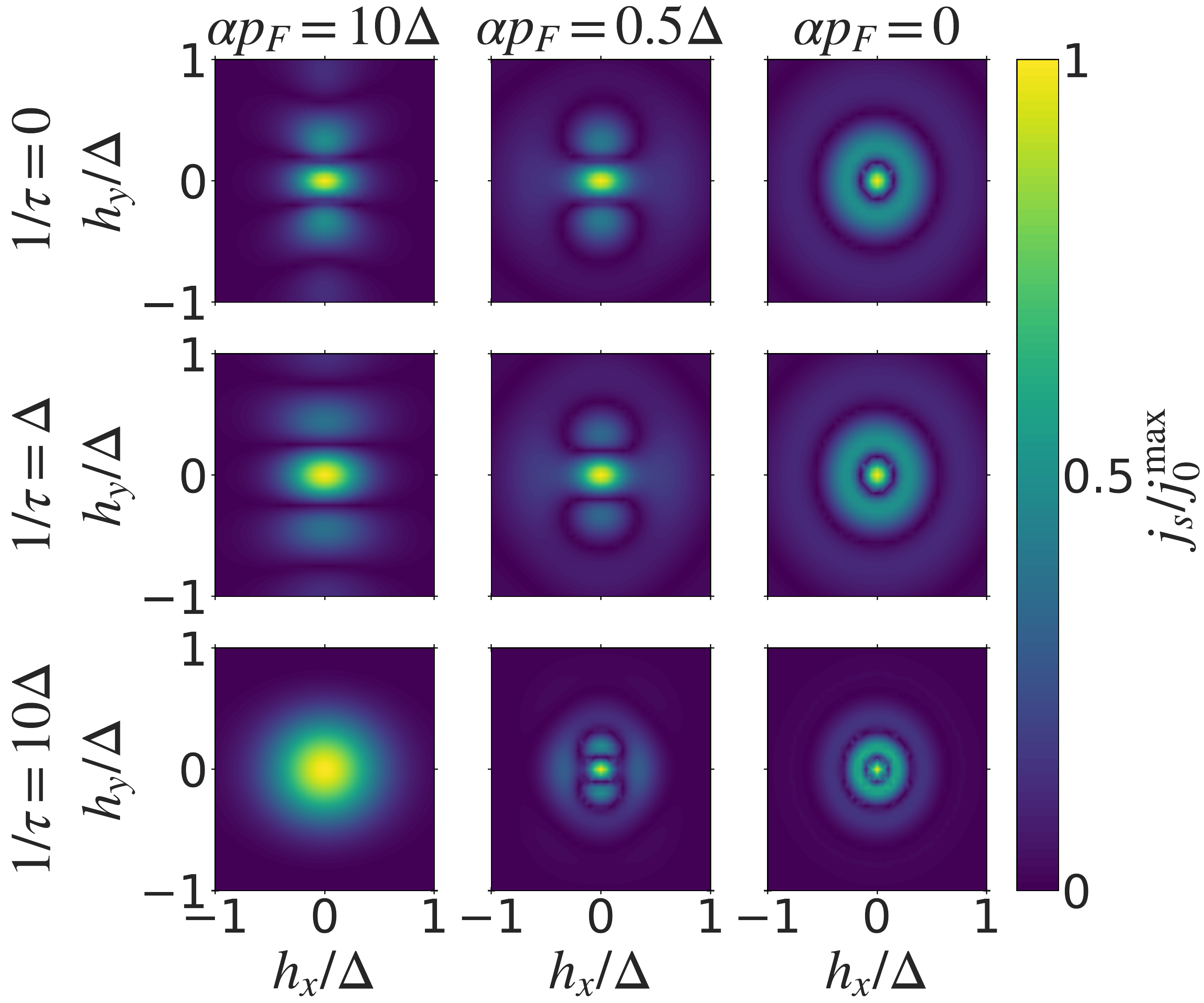} \\
    \caption{
            Normalized critical current \( j_s/j_0^{\max} \) in the \( (h_x, h_y) \) plane for pure Rashba spin–orbit coupling (\( \beta = 0 \)). 
            The junction length is fixed at \( d/\xi_0 = 2 \). 
            Rows correspond to increasing disorder strength \( 1/\tau \), while columns show increasing Rashba coupling \( \alpha p_F \). 
            In the absence of SOC, the pattern is isotropic, whereas finite Rashba coupling generates pronounced anisotropy.
            }

    \label{fig:j_hx_hy_Ea}
\end{figure}

We examine this anisotropy in more detail in Fig.~\ref{fig:j_hx_hy_Ea}, where we plot the absolute value of the critical current in the $h_x$-$h_y$ plane for various strengths of SOC and disorder. In the absence of SOC (right column), the current pattern is isotropic, whereas there is pronounced anisotropy at strong SOC and weak disorder (left column), consistent with the picture discussed above. At strong SOC coupled with strong disorder (lower-left corner of Fig.~\ref{fig:j_hx_hy_Ea}), the picture looks more isotropic again -- this is because the strong spin-relaxation in this regime suppresses all triplets. Intermediate SOC strength (middle column) produces the most intricate pattern that interpolates between the former two cases of strong SOC and no SOC.

We next consider the case of pure Dresselhaus SOC, $\alpha=0$ with $\beta\neq 0$. The corresponding current patterns in this case look the same as Fig.~\ref{fig:j_hx_hy_Ea} but rotated by $\pi/2$ in the $h_x$-$h_y$ plane. Equivalently, this corresponds to exchanging $h_x\to h_y$ and $h_y\to h_x$. 

\begin{figure}
    \centering
        \makebox[\linewidth][l]{\textbf{}}
        \includegraphics[width=\linewidth]{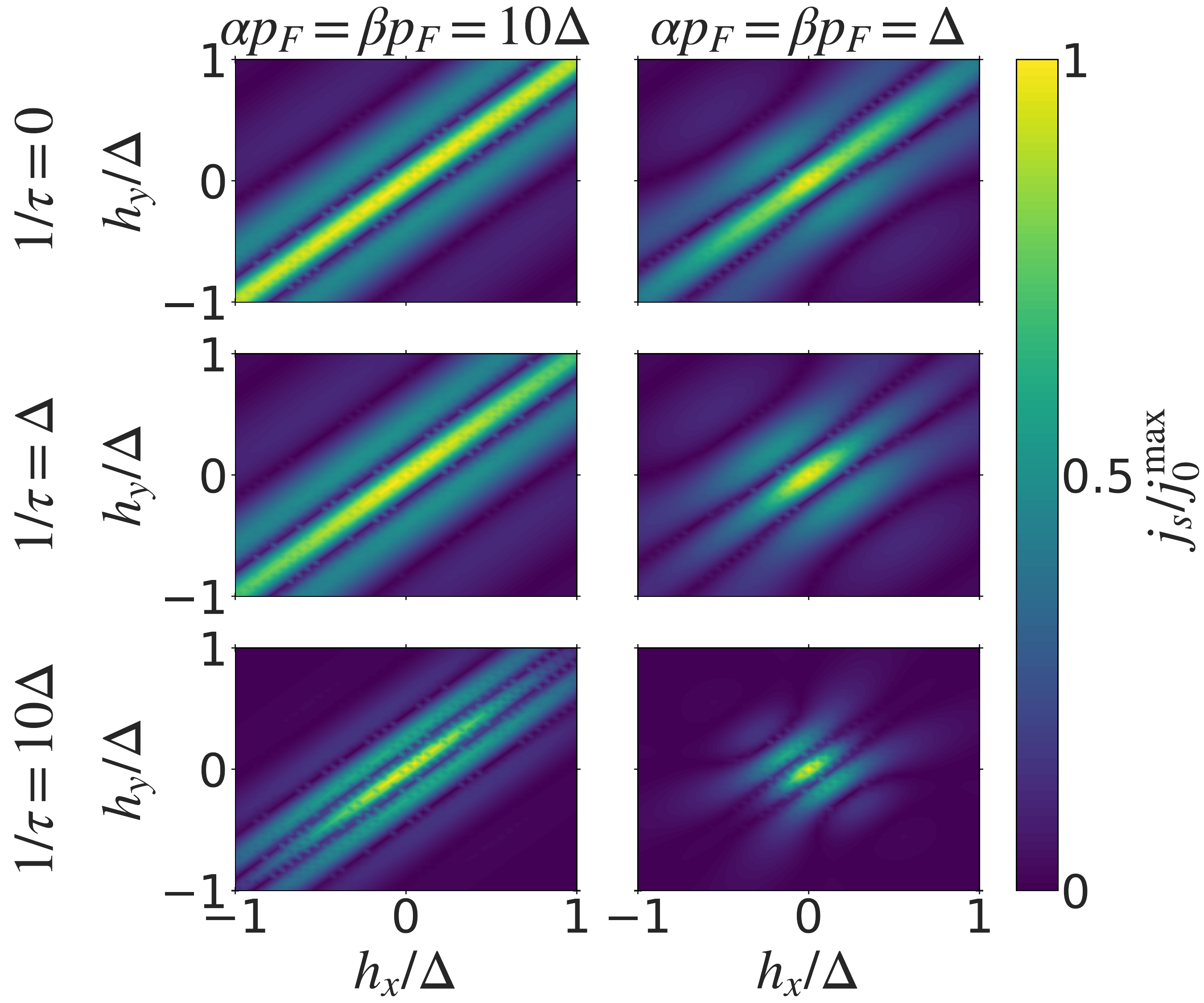} \\
    \caption{
            Normalized critical current \( j_s/j_0^{\max} \) in the \( (h_x, h_y) \) plane for equal Rashba and Dresselhaus spin–orbit coupling (\( \alpha = \beta \)). 
            The junction length is fixed at \( d/\xi_0 = 2 \). 
            Rows correspond to increasing disorder strength \( 1/\tau \), and columns to increasing SOC strength. 
            The characteristic diagonal features are signatures of the persistent spin helix, arising from the suppression of spin relaxation along the symmetry-protected direction.
            }

    \label{fig:j_hx_hy_Ea_Eb}
\end{figure}

Finally, we consider the particularly interesting case of equal Rashba and Dresselhaus SOC, $\alpha=\beta$. This regime has been widely studied in spintronics due to the emergence of the persistent spin helix \cite{bernevig2006exact}. In this situation, along the special direction $(1/\sqrt{2},1/\sqrt{2})$, the Rashba and Dresselhaus SOC exactly cancel each other, leading to the suppression of spin relaxation along that direction. In Fig.~\ref{fig:j_hx_hy_Ea_Eb} we plot the absolute value of the critical current in the $h_x$-$h_y$ plane. The distinct diagonal features visible in these plots are a direct consequence of the persistent spin helix. 

The results shown in Figs.~\ref{fig:j_hx_hy_Ea} and \ref{fig:j_hx_hy_Ea_Eb} demonstrate that different types of SOC give rise to distinct patterns of the critical current as a function of Zeeman field. The symmetry and anisotropy of these patterns are directly tied to the nature and magnitude of SOC, which could be used to experimentally probe the SOC in Josephson junctions. 

\begin{figure}
    \centering
        \includegraphics[width=\linewidth]{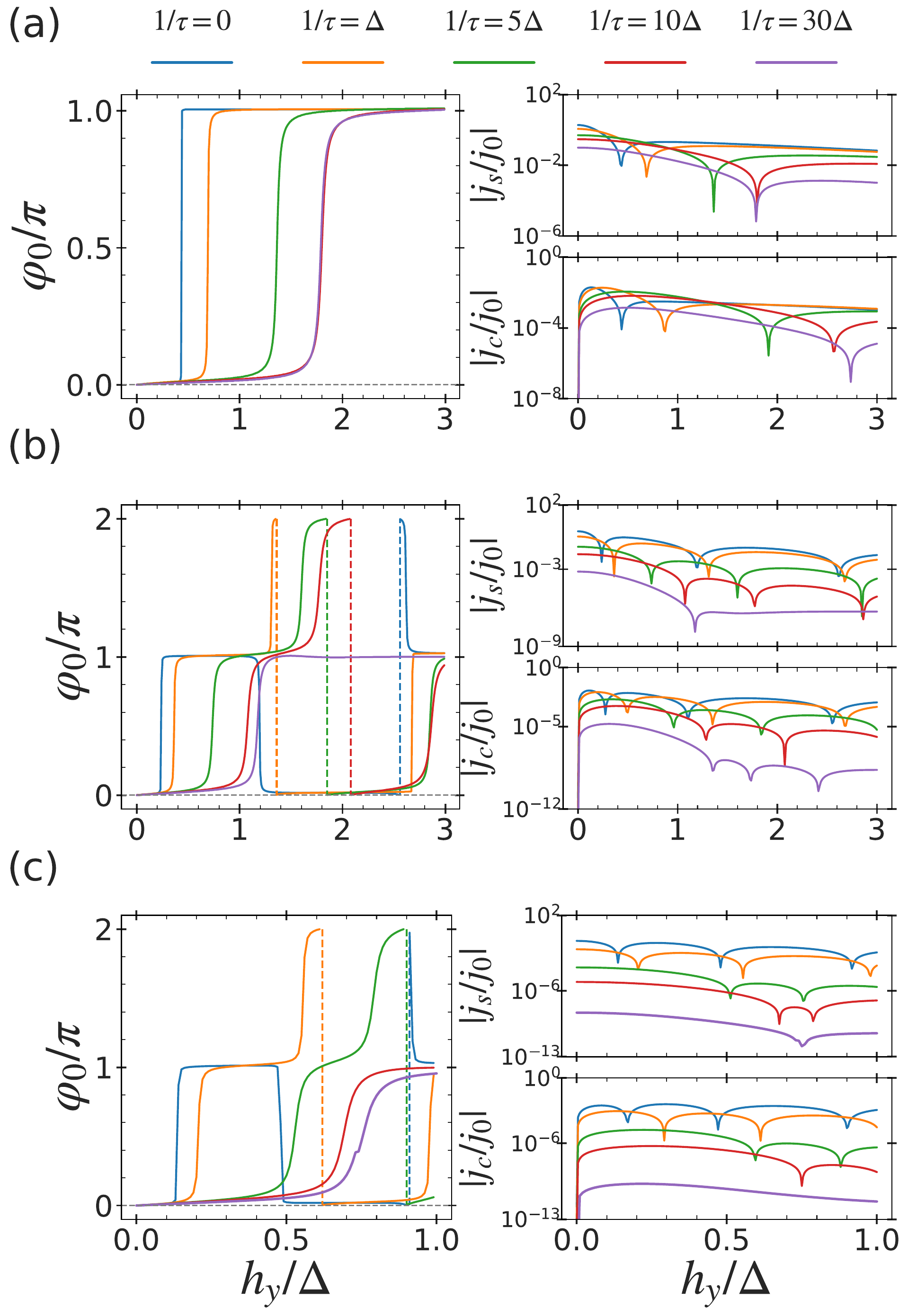}
    \caption{
            (a–c) Anomalous phase shift \( \varphi_{0} \) as a function of the in-plane Zeeman field \( h_{y} \) for junction lengths (a) \( d/\xi_0 = 0.2 \), (b) \( d/\xi_0 = 1 \), and (c) \( d/\xi_0 = 3 \). 
            The left panels show \( \varphi_{0}(h_{y}) \), while the right panels display the corresponding current components \( j_s \) and \( j_c \) (logarithmic scale), from which \( \varphi_{0} = \arctan(j_c/j_s) \) is extracted. 
            The sharp dips occur at points where either \( j_s \) or \( j_c \) changes sign. 
            Parameters: \( \alpha p_F = 10\Delta \), \( \mathcal{F}_z = 0.5\Delta \), and current normalized by \( j_0 =2 e N_0 v_F \).
          The dashed lines indicate wrapping of $\varphi_0$ from $2\pi$ to $0$. }
    \label{fig:phi_hy}
\end{figure}

\begin{figure}
    \centering
        \makebox[\linewidth][l]{\textbf{}}
        \includegraphics[width=0.7
        \linewidth]{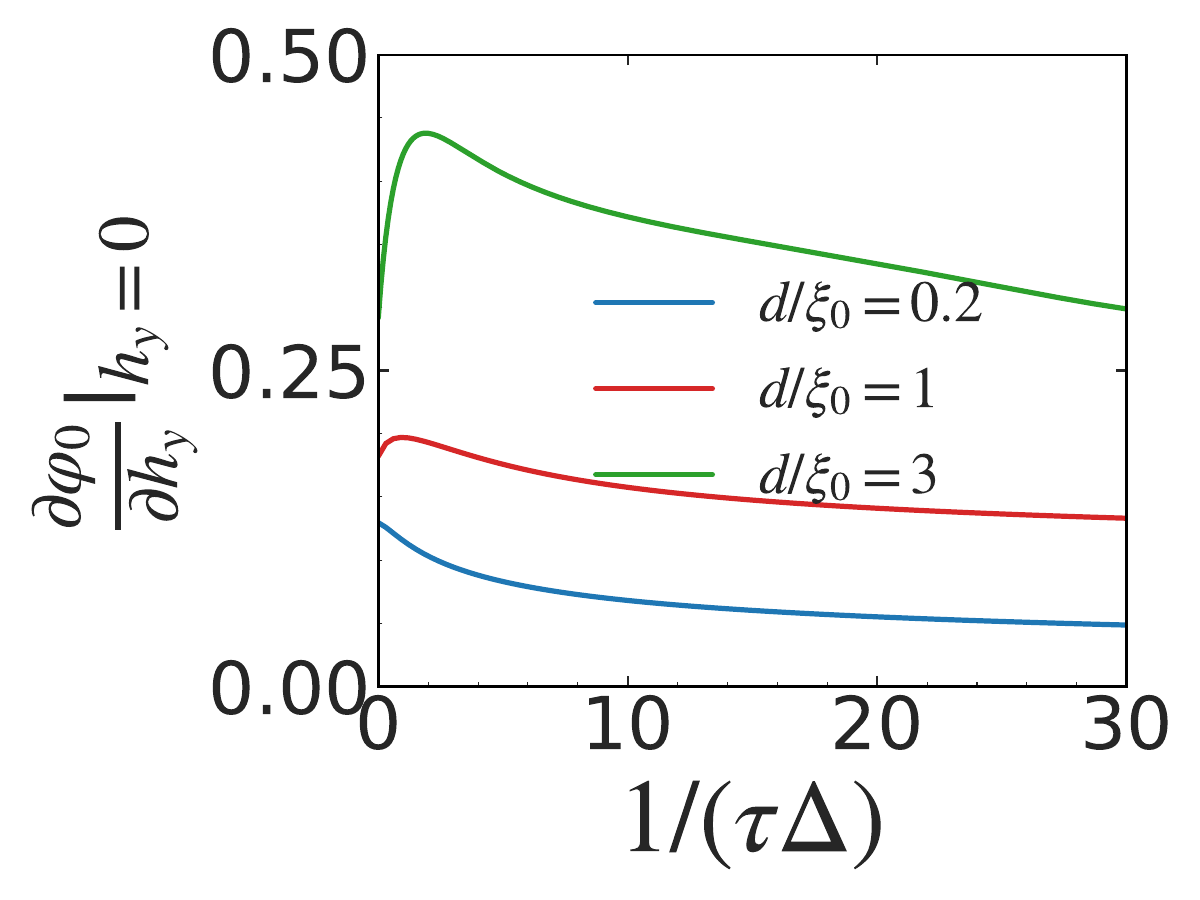} \\
    \caption{
            Slope of the anomalous phase shift at low Zeeman fields, \( \frac{\partial \varphi_0}{\partial h_y}|_{h_y=0} \), as a function of disorder strength for junction lengths \( d/\xi_0 = 0.2 \), \( 1 \), and \( 3 \). 
            The slope characterizes the linear magnetoelectric response of the junction and quantifies the sensitivity of the anomalous phase to weak in-plane fields. 
            Spin–orbit coupling parameters are the same as in Fig.~\ref{fig:phi_hy}.
            }
    \label{fig:dphi_hy_vs_Gamma}
\end{figure}

\subsection{Anomalous Josephson Effect at Arbitrary Disorder \label{SecIVb}}
Beyond modifying the conventional Josephson current $j_s$ in the presence of a Zeeman field, as discussed in Sec.~\ref{SecIVa},  SOC can give rise to a more striking phenomenon: the anomalous Josephson effect. In this case, the current-phase relation acquires a finite anomalous component $j_c$ and the anomalous phase shift $\varphi_0$ [see Eq.~\eqref{Eq:CPR}]. Within our approach, the anomalous Josephson effect originates from the term $\widehat{\mathcal{F}}$, which introduces an additional singlet-triplet conversion and enables magnetoelectric effects. We focus on the prototypical and most widely studied setup: a Rashba material subjected to an in-plane Zeeman field perpendicular to the current. Therefore, we use the following SOC and Zeeman fields in Eq.~\eqref{eq:A_matrix}
\begin{equation}
b_\mathbf{n}^x=\alpha p_F n_y, \qquad b_\mathbf{n}^y=-\alpha p_F n_x, \qquad h_\mathbf{n}^y=h_y, 
\label{Eq:coefs}
\end{equation}
and the components of the operator $\widehat{\mathcal{F}}$ defined in Eq.~\eqref{eq:F_tensor} are found as
\begin{equation}
   \mathcal{F}_x = \mathcal{F}_y = 0, \qquad \mathcal{F}_z = m\alpha^2=\frac{\alpha^2 p_F^2}{2\mu}.
\end{equation}

In Fig.~\ref{fig:phi_hy}, we plot the conventional and anomalous components of the current, $j_c$ and $j_s$, alongside the anomalous phase shift $\varphi_0$ extracted from them. The presence of magnetoelectric coupling smooths the $\varphi_0(h_y)$ dependence: in its absence, $\varphi_0$ can only have values of $0$ and $\pi$, whereas in its presence $\varphi_0$ can assume any value. We can see that the $\varphi_0$ effect persists across the entire range of disorder strengths, from the ballistic to the diffusive regime, although the overall magnitude of the current sharply decreases with increasing disorder. Moreover, we can see that, as disorder strength increases, larger fields are needed to produce a substantial $\varphi_0$ effect. This is because disorder induces triplet relaxation, and therefore stronger fields are needed to generate sufficiently large triplets to drive the $\varphi_0$ effect.

In addition, we see that in the ballistic limit, the evolution of $\varphi_0$ from $0$ to $\pi$ happens rather sharply, within a narrow range of magnetic fields, whereas in more disordered systems it is spread out over a wider field range. This might be relevant for experimental observation of this effect, since in many semiconductors with strong SOC (e.g. InAs), that may be used as a weak link, the electron $g$-factor is large \cite{rodina2025electron}, so that even a small applied field can induce a sizable Zeeman energy. In such cases it might be advantageous to work with more disordered structures, where the more gradual evolution of $\varphi_0$ over a large field range makes the effect easier to control.

To better isolate the role of disorder on the anomalous phase, we next focus on the low-field regime, where the effect scales linearly with the field, $\varphi_0\sim h_y$. In Fig.~\ref{fig:dphi_hy_vs_Gamma}, we plot the low-field slope of the $\varphi_0$ as a function of disorder strength for different junction lengths. We see that there is a strong enhancement of the effect in longer junctions, consistent with previous studies \cite{bergeret2015theory, konschelle2015theory, hasan2022anomalous}. This could be understood from the fact that SOC acts over a larger distance, allowing for greater accumulation of the anomalous phase.  Interestingly, moderate disorder leads to a slight enhancement of the $\varphi_0$ effect, particularly in longer junctions. A possible explanation for this might be that disorder makes the junction effectively longer, which promotes the $\varphi_0$ effect, as explained above. Namely, disorder reduces the superconducting coherence length $\xi$, making the ratio $d/\xi$ larger. As disorder further increases, relaxation effects become more important, leading to suppression of triplet components and gradual decline of the $\varphi_0$ effect. We note that the study of a closely related magnetoelectric phenomenon, the superconducting diode effect, has likewise reported a non-trivial enhancement of the effect at moderate disorder strengths in homogeneous Rashba superconductors \cite{hasan2025superconducting}.    

Another phenomenon related to the $\varphi_0$ effect is the superconducting Edelstein effect \cite{edelstein1995magnetoelectric,edelstein2005magnetoelectric}, whereby an applied magnetic field induces a supercurrent in a homogeneous superconductor, and inversely, a supercurrent generates a spin polarization. The Edelstein effect can also be obtained using our equations by considering a homogeneous bilayer of a superconductor in contact with a normal metal with a spin-dependent field. In our framework, this corresponds to a constant source term $\Delta(x)=\Delta$. In Appendix \ref{AppC}, we present the results obtained using our approach and show that it is consistent with the previous studies of this effect \cite{edelstein1995magnetoelectric, edelstein2005magnetoelectric, ilic2020unified}. 

The results of this section provide a useful framework for interpreting the experiments probing the $\varphi_0$ effect and related magnetoelectric phenomena. In realistic devices, the relevant parameter regime is often characterized by moderate disorder, and therefore, the simpler ballistic or diffusive theories might not be appropriate.

\section{Application to junctions with altermagnets \label{Sec:altermagnet}}

The interaction between superconductivity and altermagnetism has recently garnered significant attention \cite{sun2023andreev, papaj2023andreev, beenakker2023phase, ouassou2023dc, zhang2024finite, lu2024varphi, giil2024quasiclassical, kokkeler2025quantum}. Altermagnets have a momentum-dependent exchange field that breaks spin degeneracy even when there is no net magnetization.  Altermagnetism affects the Andreev reflection at interfaces with superconductors  \cite{sun2023andreev, papaj2023andreev, beenakker2023phase} and leads to unconventional Josephson behavior in clean junctions \cite{ouassou2023dc, zhang2024finite, lu2024varphi}. A diffusion theory for altermagnet-superconductor systems has also been recently formulated \cite{giil2024quasiclassical, kokkeler2025quantum}, and applied to diffusive Josephson junctions \cite{heras2025interplay}. 

As discussed in the Introduction, despite the absence of net magnetization, altermagnet-superconductor structures remarkably exhibit many phenomena characteristic of conventional superconductor-ferromagnet systems,  including FFLO-like states and $0$-$\pi$ transitions. At the same time, altermagnets may offer practical advantages over conventional ferromagnets, since the lack of stray fields in altermagnets \cite{vsmejkal2022emerging} means that detrimental orbital effects do not arise. Here we investigate the Josephson junctions with altermagnets across the full range of disorder strength, generalizing previous studies for clean \cite{ouassou2023dc, zhang2024finite, lu2024varphi} and diffusive \cite{heras2025interplay} junctions. We demonstrate that even a small amount of disorder is sufficient to wash out features such as $0$-$\pi$ transitions and the anisotropic critical current. 

As a minimal description of an altermagnet, we assume the $d$-wave momentum dependence  with magnetization axis along the $z$ direction, captured by the following term in Eq.~\eqref{eq:A_matrix}
\begin{equation}
    h_\mathbf{n}^z = h_0 \left[ (n_x^2 - n_y^2)\cos(2\chi) 
    + 2 n_x n_y \sin(2\chi) \right].
\end{equation}
Here, $h_0$ sets the amplitude of altermagnetism, and $\chi$ defines the orientation of the $d$-wave pattern with respect to the current direction. 

Following Sec.~\ref{SecIIIb}, we find the critical current analytically in the ballistic limit 
\begin{equation}
j_s^B=  e N_{0}v_{F} \pi T \Delta^2 \sum_{\omega_n>0,\pm}\left\langle  \frac{|n_x|}{ (\omega_n\pm i h_\mathbf{n}^z)^2} e^{-\frac{2 d (\omega_n\pm i h_\mathbf{n}^z)}{v_F|n_x|}} \right\rangle ,
\end{equation}
and in the diffusive limit
\begin{equation}
j_s^D=e N_{0}\pi T D\sum_{\omega_n>0,\pm}\frac{\Delta^2 \kappa_{\omega}^\pm e^{-d \kappa_{\omega}^\pm}}{(\omega_n+\Gamma)^2},
\label{eq:Diffusive_Altm}
\end{equation}
where we introduced 
\begin{equation}
\kappa_\omega^\pm=\sqrt{\frac{2(\omega_n+\Gamma)}{D (1\pm i K_\chi)}}, \quad 
\Gamma=h_0^2\tau, \quad K_\chi=3h_0\tau   \cos 2\chi.
\label{Eq:parameters}
\end{equation}
In the present representation, the altermagnet in the diffusive limit is governed by two parameters: the relaxation rate $\Gamma$ that has a form similar to the Dyakonov-Perel mechanism \cite{dyakonov1972spin}, and the parameter $K_\chi$, which accounts for spin-dependent corrections to the diffusion constant (see also Sec.~\ref{SecIIIb}). Retaining $K_\chi$ is necessary to obtain the $0$-$\pi$ transitions and the directional dependence of the critical current. Specifically, $K_\chi$ introduces an imaginary component to the decay rates $\kappa_\pm$, which can generate the oscillatory behavior of the current and sign changes. However, these contributions arise from higher-order terms in the diffusive limit ($DK_\chi\sim \tau^2$) and are therefore weak. On the other hand, the relaxation term $\Gamma$ is typically larger ($\Gamma\sim\tau$). Its primary effect is to suppress the overall magnitude of the current, and also to damp the oscillatory features.  The expressions for $\Gamma$ and $K$ presented in \eqref{Eq:parameters} agree with Ref.~\cite{heras2025interplay}, where these parameters have been derived using other methods.

\begin{figure}
    \centering
        \makebox[\linewidth][l]{\textbf{}}
        \includegraphics[width=\linewidth]{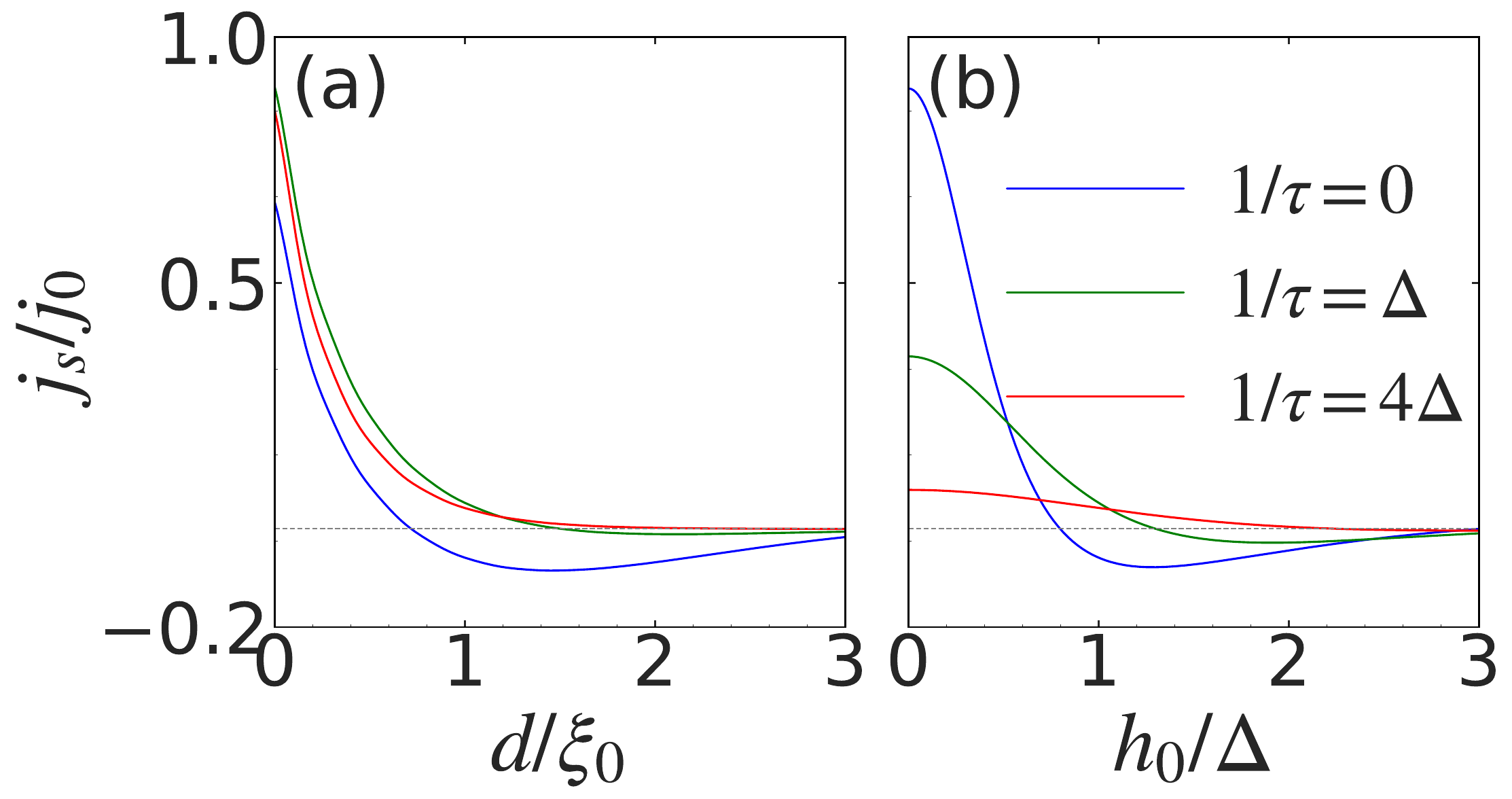} \\
    \caption{
            Josephson critical current in an altermagnetic SNS junction at $\chi=0$. (a) Dependence on junction length $d$ at fixed $h_0=\Delta$. 
            (b) Dependence on exchange field strength $h_0$ at fixed $d/\xi_0=\Delta$. Currents are normalized to $j_0 =2 e N_0 v_F$.
            }
    \label{fig:j_d_h0_ALM}
\end{figure}

\begin{figure}[ht]
    \centering
    \includegraphics[width=0.7\linewidth]{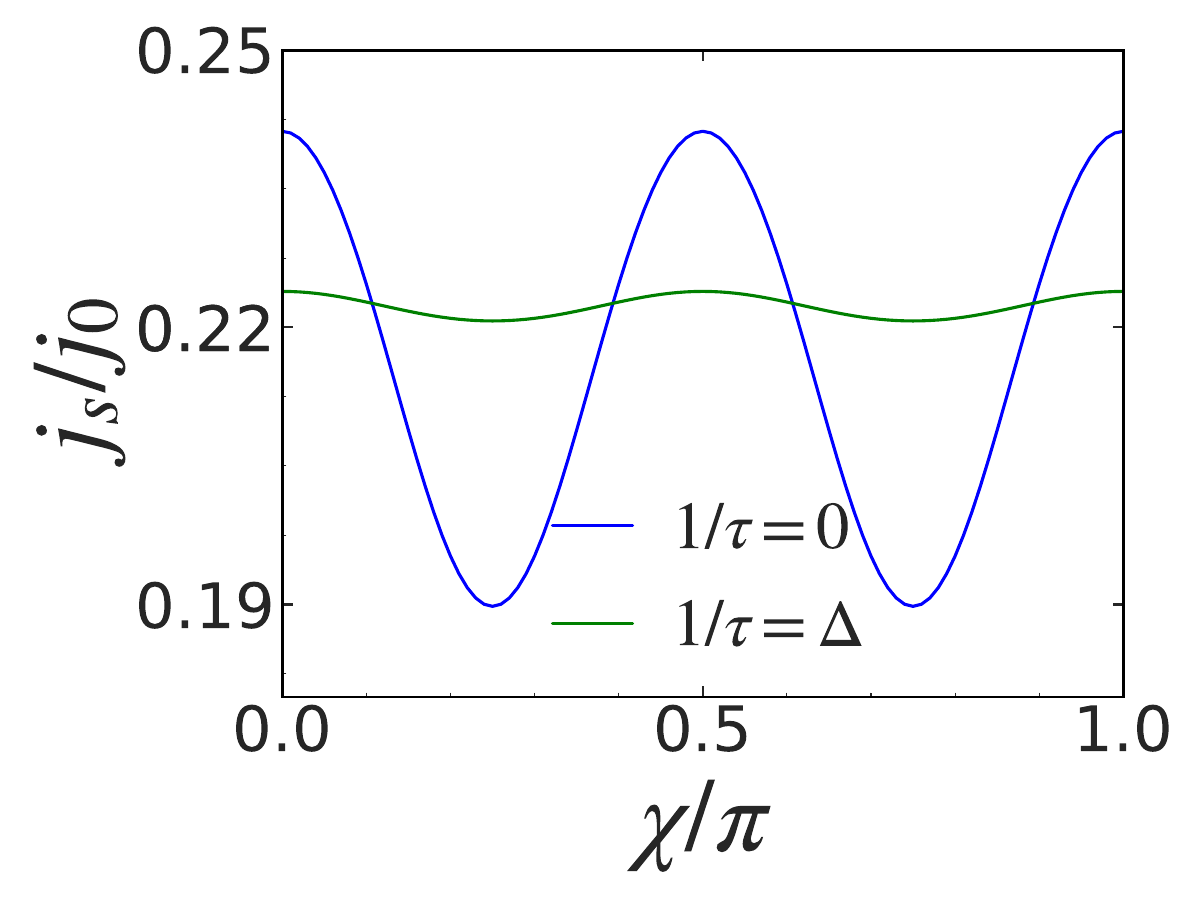}
        \caption{
                Josephson critical current in an altermagnet junction as a function of the angle $\chi$. 
                We set $h_0 = 0.5\Delta$ and $d/\xi_0 = \Delta$. 
                Currents are normalized to $j_0 =2 e N_0 v_F$. 
                }

    \label{fig:j_chi_ALM}
\end{figure}

We plot the Josephson critical current $j_s$ for the altermagnetic Josephson junction in Fig.~\ref{fig:j_d_h0_ALM} as a function of junction length $d$ and altermagnet strength $h_0$, at different strengths of disorder. In the ballistic limit, a clear $0$-$\pi$ transition appears when varying either $d$ or $h_0$. However, once weak disorder is introduced ($\tau^{-1}\gtrsim \Delta$), the transition becomes barely discernible, although it can still occur. This is consistent with our analysis in the diffusive limit above.  In Fig.~\ref{fig:j_chi_ALM}, we show another feature of the altermagnet -- the direction dependence of the critical current. Namely, its magnitude depends on the orientation of the altermagnetic $d$-wave "flower" pattern relative to the current, captured by the angle $\chi$. The amplitude of the oscillations with respect to $\chi$ is also suppressed by a small amount of disorder. 

The results of this section indicate that very clean samples would likely be required to clearly observe the $0$-$\pi$ transition or directional dependence of the critical current in experiments. We emphasize, however, that our theory is limited to weak altermagnets,  where the Fermi energy greatly exceeds the altermagnetic splitting, and the quasiclassical approximation holds ($\mu\gg h_0$).

\section{Conclusion \label{SecVI}}

We have presented a comprehensive theory of the Josephson effect in SNS junctions containing spin-dependent fields at arbitrary disorder strength. In Sec.~\ref{SecIII}, by solving the linearized quasiclassical Eilenberger equation, we derived a general expression for the Josephson current that is valid beyond the conventional ballistic and diffusive limit. 

Using this approach, we analyzed several experimentally relevant situations. In junctions with Rashba and Dresselhaus SOC, we examined how the conventional Josephson current evolves in an applied magnetic field, and discussed how this evolution can be used to experimentally probe the type and strength of SOC in the junction (\ref{SecIVa}). We also investigated the anomalous  Josephson effect in systems with Rashba SOC in Sec.~\ref{SecIVb}, and found that the anomalous phase shift $\varphi_0$ remains robust over the broad range of disorder strength. In sufficiently long junctions, moderate disorder can even enhance $\varphi_0$, highlighting the nontrivial interplay between SOC and impurity scattering. Finally, we studied the Josephson effect in altermagnetic materials in Sec.~\ref{Sec:altermagnet},  and demonstrated that the $0$-$\pi$ transition present in the clean systems is rapidly suppressed by disorder.

Although here we have mostly focused on the Josephson effect, the method we used can be readily extended to other situations. For instance, as demonstrated in Appendix ~\ref{AppC}, it can be adapted straightforwardly to homogeneous bilayer systems. More generally, this approach can be applied to a wide class of semi-infinite hybrid systems that can be treated with Fourier transforms. Extensions to unconventional superconducting pairing are also possible by introducing the appropriate source terms. In addition to the supercurrent, the solution presented in Eq. ~\eqref{eq:full_solution} can also be used to compute other observables, such as induced magnetic moments or corrections to the normal-state density of states due to superconductivity.

\begin{acknowledgments}
 This work is part of the Finnish Centre of Excellence in Quantum Materials (QMAT). We acknowledge the financial support of the Finnish Ministry of Education and Culture through the Quantum Doctoral Education Pilot Program (QDOC VN/3137/2024-OKM-4) and the Research Council of Finland through the Finnish Quantum Flagship project 359240. S.I. is supported by the Research Council of Finland (Grant Number 355056). 
 F.~S.~B. thanks financial support from  the Spanish MCIN/AEI/10.13039/501100011033 through the grant PID2023-148225NB-C31, and the
European Union’s Horizon Europe research and innovation program under grant agreement No. 101130224 (JOSEPHINE).
\end{acknowledgments}

\appendix 
\section{Analytical results for ballistic Josephson junctions with Rashba SOC \label{AppA}}

In Sec.~\ref{SecIIIb}, we discussed how the general solution \eqref{eq:full_solution} simplifies in the ballistic limit and illustrated the calculation of the Josephson current for the simplest case without spin-dependent fields. Here, we follow the same strategy for a junction with Rashba spin--orbit coupling (SOC) and an in-plane Zeeman field, described by the coefficients introduced in Eqs.~\eqref{Eq:coefb} and \eqref{Eq:coefh}.
For simplicity, in the following we neglect the magnetoelectric term $\widehat{\mathcal{F}}$. While an analytical solution can, in principle, also be obtained in its presence, the resulting expressions become rather cumbersome and are not presented here.

Starting from the ballistic-limit solution, Eq.~\eqref{eq:ballistic}, after Fourier-transforming it to the real space, the anomalous Green's function in the middle of the junction, $\hat{f}(0)=f^i(0) \sigma_i$ ($i=0,x,y,z$), can be written as
\begin{equation}
f^{i}(0) =
\sum_{\alpha=1}^{4}
a_\alpha^{i}
\left(
f_\alpha^{c}\cos\frac{\varphi}{2}
+
f_\alpha^{s}\sin\frac{\varphi}{2}
\right),
\qquad
\label{eq:ballistic_limit}
\end{equation}
where $\varphi$ is the superconducting phase difference and
\begin{equation}
f^{c}_{\alpha}=-\frac{i\Delta}{\kappa_\alpha}
e^{-\frac{d\kappa_\alpha}{v_F|n_x|}},
\qquad
f^{s}_{i}=\mathrm{sgn}(n_x)\frac{\Delta}{\kappa_\alpha}
e^{-\frac{d\kappa_\alpha}{v_F|n_x|}} .
\end{equation}
The four decay rates $\kappa_\alpha$ are
\begin{equation}
\kappa_{1,2}=\omega\pm\chi_{+},\qquad
\kappa_{3,4}=\omega\pm\chi_{-},
\end{equation}
where
\begin{align}
\chi_{\pm}&=\frac{1}{\sqrt{2}}
\sqrt{\pm M-h^{2}-E_\alpha^{2}}, \nonumber\\
M&=\sqrt{h^4+E_\alpha^4+2h^2E_\alpha^2\cos2(\theta-\rho)},
\end{align}
and $E_\alpha=\alpha p_F$ is the Rashba energy. We introduced the angle $\rho$ between the current and the field directions, so that $\tan\rho=h_x/h_y$. The corresponding weights $a_\alpha^i$ are
\begin{equation}
\begin{aligned}
    &a^{0}_{1}=a^{0}_{2}=\frac{1}{4}+\frac{E_\alpha^2-h^2}{4M}, \quad a^{0}_{3}=a^{0}_{4}=\frac{1}{4}-\frac{E_\alpha^2-h^2}{4M}, \quad \\
   &a^{x}_{1}=-a^{x}_{2}=-\frac{ih[E_\alpha^{2}\cos(2\theta- \rho)+(h^2-M)\cos\rho]}{4\chi_{+} M}, \\
   &a^{x}_{3}=-a^{x}_{4}=-\frac{ih[E_\alpha^{2}\cos(2\theta- \rho)+(h^2+M)\cos\rho]}{4\chi_{-} M}, \\
   &a^{y}_{1}=-a^{y}_{2}=\frac{ih[E_\alpha^{2}\sin(2\theta- \rho)+(h^2-M)\sin\rho]}{4\chi_{+} M}, \\
   &a^{y}_{3}=-a^{y}_{4}=-\frac{ih[E_\alpha^{2}\sin(2\theta- \rho)+(h^2+M)\sin\rho]}{4\chi_{+} M}  \\
   &a^{z}_{1}=a^{z}_{2}=-a^z_3=-a_4^z=\frac{i h E_\alpha \cos[\theta - \rho]}{2 M}.\end{aligned}
\end{equation}

The analytical expression for the Josephson current is obtained by substituting the expression \eqref{eq:ballistic_limit} into Eq.~\eqref{eq:current}. This expression reproduces the numerical results shown in Fig.~\ref{fig:j_hx_hy_Ea} in the ballistic limit.

\section{Diffusion operator in systems with Rashba SOC \label{AppB}}
In Eq.~\eqref{eq:diffusion}, we derived the general form of the diffusion operator for generic spin-dependent fields. Here, we present its explicit form for the case of Rashba SOC and in-plane Zeeman field, described by the coefficients introduced in Eqs.~\eqref{Eq:coefb} and \eqref{Eq:coefh}. Substituting these coefficients into Eq.~\eqref{eq:diffusion}, we obtain

\begin{widetext}
\begin{equation}
\widehat{\mathcal{D}}^{-1} =(2\omega +D Q^{2})+ \begin{pmatrix}
\Gamma_{R} &0 & -2 i \sqrt{D\Gamma_R} Q& -2 i h_x \\
0 & \Gamma_{R}  & 0 & -i(2h_y-\Gamma_{st} Q) \\
 -2i\sqrt{D\Gamma_R}Q & 0 & 2 \Gamma_{R} & 0 \\
-2 i h_x &  -i(2 h_y-\Gamma_{st} Q) & 0 & 0
\end{pmatrix},
\label{eq:Dif}
\end{equation}
\end{widetext}
here $D=\frac{1}{2}v_{F}^{2}\tau$ is the diffusion constant, $ \Gamma_R=2(\alpha p_F)^2\tau$ is the D’yakonov--Perel spin-relaxation rate, and $\Gamma_{st}=(\alpha p_F)^3\tau^2/p_F$
characterizes the singlet--triplet coupling induced by the spin--orbit interaction. Eq.~\eqref{eq:Dif} is the familiar singlet-triplet (spin-charge) coupled diffusion operator widely used in superconducting and normal-state transport \cite{burkov2004theory}.

\section{Edelstein effect \label{AppC}}

In this Appendix, we calculate the inverse Edelstein effect: the anomalous current generated by applying a Zeeman field in a superconducting system with SOC. Our goal here is primarily to show that our approach can be readily extended to homogeneous bilayers and other geometries, beyond the Josephson junctions considered in the main text. 

\begin{figure}[htbp]
    \includegraphics[width=0.7\linewidth]{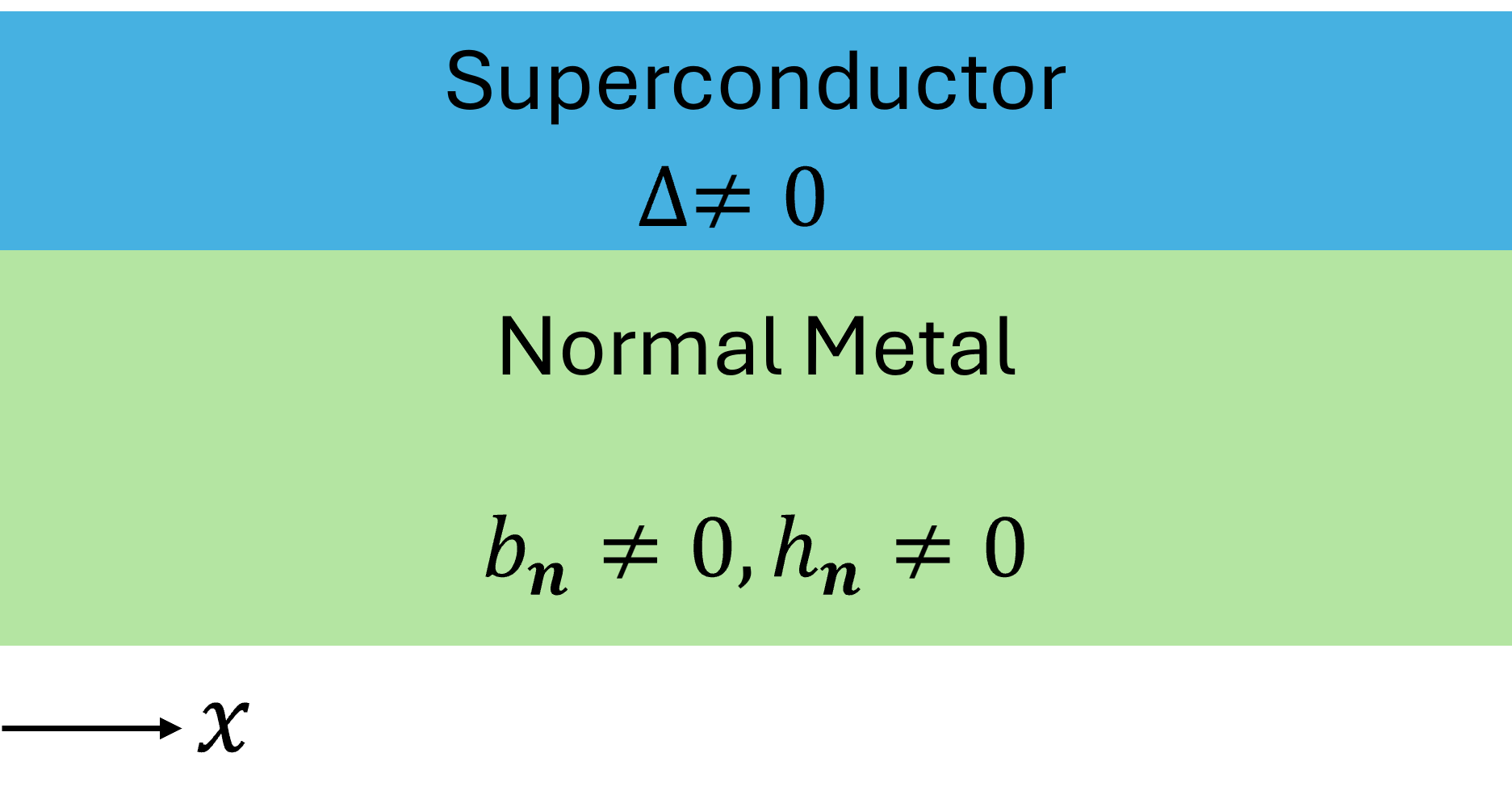}
    \caption{Homogeneous bilayer of a normal metal with Rashba SOC and an in-plane Zeeman field, and a superconductor.}
    \label{bulkmodel}
\end{figure}

\begin{figure}[htbp]
    \centering
    \includegraphics[width=0.8\linewidth]{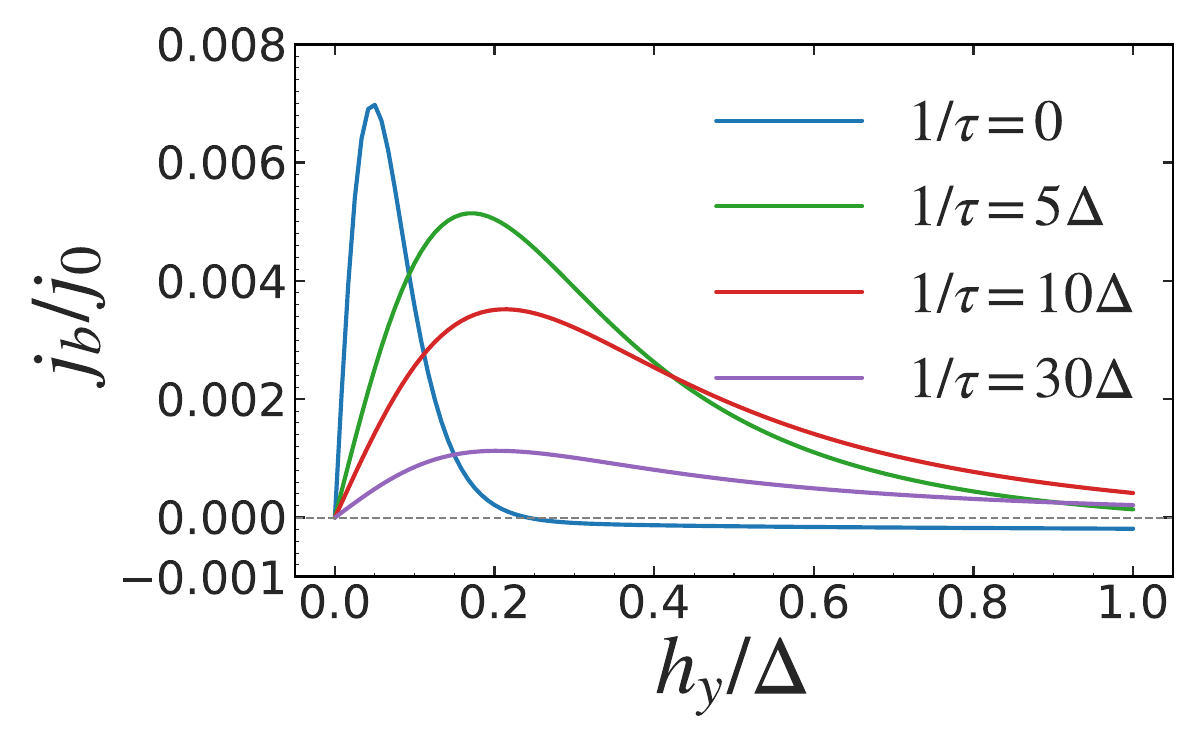}
    \caption{Bulk anomalous current $j_b$ versus in-plane Zeeman field $h_y$ in a Rashba spin–orbit coupled normal metal proximitized by a superconductor. Currents are normalized by $j_0 =2 e N_0 v_F$.
}
    \label{Fig:Edelstein}
\end{figure}

We consider a homogeneous bilayer system, shown in Fig.~\ref{bulkmodel}, consisting of a superconductor and a normal metal with finite Rashba SOC and an in-plane Zeeman field in the $y$-direction, described by the coefficients introduced in Eq.~\eqref{Eq:coefs}. 
To model the bilayer,  we set $\Delta_Q=-2i \Delta \sqrt{2\pi}\delta(Q)$, corresponding to the homogeneous pair potential.  Then, we simply have $\hat{f}=\hat{f}_Q(Q=0)$ since the system is homogeneous, and the current follows from Eq.~\eqref{eq:current}. We plot the anomalous current along the $x$-direction obtained this way in Fig.~\ref{Fig:Edelstein}. This result generalizes the previous studies of the Edelstein effect beyond the weak-field regime \cite{edelstein1995magnetoelectric, edelstein2005magnetoelectric, ilic2020unified}.

Next, we investigate the low-field regime by expanding the current up to first order in $h$. There, the anomalous current can be found analytically in the form $j_b=\chi_E h_y$, where $\chi_E$ is the Edelstein response function
\begin{multline}
\chi_E=e \pi N_0 v_F T  \frac{(\alpha p_F)^3\Delta^2}{8\mu} \times \\\sum_{\omega_n>0} \frac{1}{\omega_n^2}\frac{1}{ \omega_n(2\omega_n+1/\tau)^2+\alpha^2p_F^2(4\omega_n+1/\tau)}.
\end{multline}
This result is identical to the Edelstein response function obtained in the study of the direct Edelstein effect
\cite{ilic2020unified}, as expected from Onsager reciprocity. 

\bibliography{references}  

\end{document}